\documentclass[10pt,english]{article}
\usepackage{times}
\usepackage[T1]{fontenc}
\usepackage[latin1]{inputenc}
\usepackage{amsmath}
\usepackage{amssymb}

\makeatletter

\newcommand{\noun}[1]{\textsc{#1}}
\providecommand{\tabularnewline}{\\}

\usepackage{slashed}

\usepackage{babel}
\makeatother
\begin{document}

\title{\textbf{NEUTRINO MASS AS A CONSEQUENCE OF THE EXACT SOLUTION OF 3-3-1
GAUGE MODELS WITHOUT EXOTIC ELECTRIC CHARGES}}

\author{\noun{ADRIAN} PALCU}

\date{\emph{Department of Theoretical and Computational Physics - West
University of Timi\c{s}oara, V. P\^{a}rvan Ave. 4, RO - 300223 Romania}}

\maketitle
\begin{abstract}
The unjustly neglected method of exactly solving generalized electro-weak
models - with an original spontaneous symmetry breaking mechanism
based on the gauge group $SU(n)_{L}\otimes U(1)_{Y}$ - is applied
here to a particular class of chiral 3-3-1 models. This procedure
enables us - without resorting to any approximation - to express the
boson mass spectrum and charges of the particles involved therein
as a straightforward consequence of both a proper parametrization
of the Higgs sector and a new generalized Weinberg transformation.
We prove that the resulting values can be accommodated to the experimental
ones just by tuning a sole parameter. Furthermore, if we take into
consideration both the left-handed and right-handed components of
the neutrino (included in a lepton triplet along with their corresponding
left-handed charged partner) then we are in the position to propose
an original method for the neutrino to acquire a very small but non-zero
mass without spoiling the previously achieved results in the exact
solution of the model. In order to be compatible with the existing
phenomenological data, the range of that sole parameter imposes in
our method a large order of magnitude for the VEV $<\phi>\sim10^{6}$
TeV. Consequently, the new bosons of the model have to be very massive.


\end{abstract}

\section{Introduction}

In Ref. \cite{key-1} devoted to exactly solving generalized chiral
electro-weak $SU(n)_{L}\otimes U(1)_{Y}$ gauge models, a new Higgs
mechanism has been proposed that acts as a good mathematical strategy
able to break the symmetry of the model up to the universal residual
one $U(1)_{em}$, as in the Standard Model (SM). A non-degenerated
boson mass spectrum is therefore expected once the spontaneous symmetry
breaking (SSB) is accomplished. This could be achieved by introducing
an adequate metric into the kinetic term of the scalar sector of the
model. Then, based on a new and generalized Weinberg transformation
(gWt), we can separate the neutral gauge fields keeping at the same
time massless the electromagnetic one. Furthermore, one can give the
particles their charges (both electric and neutral ones) by identifying
the coupling coefficients of certain currents in the general method.
Fermion masses arise from special Yukawa couplings consisting of tensor
products among certain Higgs multiplets that get the usual form of
Yukawa couplings when boosted towards unitary gauge. 

For these models the renormalizability criteria are also satisfied: 

\begin{itemize}
\item \emph{The axial anomaly cancellation} - that can in turn explain the
number of fermionic generations in concrete models - has a specific
formulation (see Sec. 6.2. in Ref. \cite{key-1}).
\item \emph{The formal dimension of terms in the new Yukawa couplings} -
by means of which one controls the divergences that could appear in
the propagator approach - is maintained at a suitable level (see Sec.
4.3. in Ref \cite{key-1}).
\end{itemize}
Reference \cite{key-1} ends up with the successful confirmation of
the general method in the particular case of the Pisano, Pleitez and
Frampton (PPF) model. Designed to explain the electro-weak interaction
based on the local gauge group $SU(3)_{c}\otimes SU(3)_{L}\otimes U(1)_{Y}$,
this was the first of the so called ''3-3-1 models'' but not the
only one. The PPF model (Refs. \cite{key-2,key-3}) had to pay the
price of dealing with exotic electric charges for some quarks ($\pm5e/3,$
and $\pm4e/3$ respectively) as doubly charged bileptons were allowed
to occur in the model. Concerning the gauge sector, it is obvious
that for algebrical reasons a new neutral physical gauge boson (besides
the one coming from SM) has to occur. Due to it, a new neutral charge
is added to the ''old'' charges from SM in order to fully describe
the fermion particles in the model. In order to improve the particle
content of 3-3-1 models, new family representations were proposed
as candidates for the fermion sector. Therefore, some new 3-3-1 models
that avoid exotic electric charges have emerged in the literature
\cite{key-4} - \cite{key-8}. Regarding the manner in which those
models accomplish the anomaly cancellation requirement, they can be
either ''family models'' (with generations treated differently)
or ''one-family models'' (with generations obtained by replication).
There are some papers \cite{key-9} - \cite{key-13} that carry a
systematic analysis of those possible models, keeping alive the anomaly
cancellation principle in each particular case and even predicting
new kinds of models by combining the anomaly factors so that they
vanish by an interplay. 

In the following, we assume the results of the general method in Ref.
\cite{key-1} as valuable and useable in concrete cases and apply
them to a particular class of ''no-exotic-electric-charge'' 3-3-1
models (namely, models C and D in Ref. \cite{key-9}) attempting to
find its exact solution. Finally, the correlation with experimental
data (Ref. \cite{key-14}) can be made just by tuning certain parameters
of the model. At the same time we show that this procedure allows
us to propose a mechanism for the neutrino to acquire a very small
- \emph{but non-zero!} - Majorana mass when the breaking scale of
the model reaches a very large order of magnitude.

The paper is organized as follows. The main results of the general
method are briefly presented in Section 2, as they stand as the starting
point to our results. Section 3 deals with the exact solution of a
particular class of 3-3-1 models. All the SM phenomenology is reproduced.
The boson masses and the charges of the particles are obtained just
by making an appropriate parameter choice in the general method. This
leads straightforwardly to the spinor sector's content of the models
C and D from Ref. \cite{key-9}. Then we present the Yukawa terms
that enable fermion families to acquire their masses. Section 4 focuses
on the neutrino mass issue by introducing a symmetric matrix with
sextet transformation involving tensor products of Higgs multiplets.
An appropriate way to embed neutrino mixing in our method is also
shown with special emphasis on its compatibility with experimentally
obtained mass squared differences. In the concluding remarks the resulting
neutrino mass range and its implications for the exact solution of
3-3-1 model are discussed.

\section{Preliminaries}

Let us review the main results of exactly solving the generalized
$SU(n)_{L}\otimes U(1)_{Y}$ gauge model as they come from Ref. \cite{key-1}.
In accordance with the general method, the scalar sector of any ''pure
left'' gauge model must consist of $n$ Higgs multiplets $\phi^{(1)}$,
$\phi^{(2)}$, ... , $\phi^{(n)}$ satisfying the orthogonal condition
$\phi^{(i)+}\phi^{(j)}=\phi^{2}\delta_{ij}$ in order to eliminate
unwanted Goldstone bosons after SSB. Here $\phi$ is a gauge-invariant
real field variable and $n$ is the dimension of the fundamental irreducible
representation of the gauge group. The parameter matrix $\eta=\left(\eta_{0},\eta{}^{(1)},\eta{}^{(2)},...,\eta{}^{(n)}\right)$
with the property $Tr\eta^{2}=1-\eta_{0}^{2}$ is also introduced.
Then, the Higgs Lagrangean density (Ld) stands:

\begin{equation}
\mathcal{L}_{H}=\frac{1}{2}\eta_{0}^{2}\partial_{\mu}\phi\partial^{\mu}\phi+\frac{1}{2}\sum_{i=1}^{n}\left(\eta{}^{(i)}\right)^{2}\left(D_{\mu}\phi^{(i)}\right)^{+}\left(D^{\mu}\phi^{(i)}\right)-V(\phi)\label{Eq.1}\end{equation}
where $D_{\mu}\phi^{(i)}=\partial_{\mu}\phi^{(i)}-ig(A_{\mu}+y^{(i)}A_{\mu}^{0})\phi^{(i)}$
means covariant derivatives of the model. After the SSB the boson
masses take the forms: \begin{equation}
M_{i}^{j}=\frac{1}{2}g\left\langle \phi\right\rangle \sqrt{\left[\left(\eta^{(i)}\right)^{2}+\left(\eta^{(j)}\right)^{2}\right]}\label{Eq.2}\end{equation}
for the non-diagonal gauge bosons (which usually are charged but,
as we can easily observe in the 3-3-1 models under consideration in
the next section, one of them could also be neutral) and:

\begin{equation}
\left(M^{2}\right)_{ij}=\left\langle \phi\right\rangle ^{2}Tr\left(B_{i}B_{j}\right)\label{Eq.3}\end{equation}
with: \begin{equation}
B_{i}=g\left[D_{i}+\nu_{i}\left(D\nu\right)\frac{1-\cos\theta}{\cos\theta}\right]\eta\label{Eq.4}\end{equation}
for the diagonal gauge bosons of the model. The last ones are neutral
without exception. The $\eta$ parameter diagonal matrix comes from
the scalar sector and it will essentially determine the mass spectrum
of the model, $\theta$ is the rotation angle around the versor $\nu$
orthogonal to the electromagnetic direction in the parameter space
\cite{key-1}. The versor condition holds $\nu_{i}\nu^{i}=1.$ For
the concrete models we will work on, $D$s are the Hermitian diagonal
generators (Cartan subalgebra) of the $SU(3)_{L}$ group, \emph{i.e.}
$D_{1}=T_{3}$ and $D_{2}=T_{8}$ connected to the Gell-Mann matrices
in the manner $T_{a}=\lambda_{a}/2$. 

After we extracted the electromagnetic potential (the field $A_{\mu}^{em}$
that is massless), a special $SO(n-1)$ transfomation $\omega$ remains
to be determined in each particular case, in order to bring the mass
matrix into the physical basis $(A_{\mu}^{em},Z_{\mu},Z_{\mu}^{\prime})$
so that the masses of physical neutral bosons are just the eigenvalues
of the diagonal form of the matrix (3). The gWt can be defined as: 

\begin{equation}
A_{\mu}^{0}=A_{\mu}^{em}\cos\theta-\nu_{i}\omega_{\cdot j}^{i\cdot}Z_{\mu}^{j}\sin\theta\label{Eq.5}\end{equation}

\begin{equation}
A_{\mu}^{k}=\nu^{k}A_{\mu}^{em}\sin\theta+\left[\delta_{i}^{k}-\nu^{k}\nu_{i}\left(1-\cos\theta\right)\right]\omega_{\cdot j}^{i\cdot}Z_{\mu}^{j}\label{Eq.6}\end{equation}

Moreover, the charges of the particles can be identified as coupling
coefficients of the currents. In a certain representation $\rho$
the charge operators have the following diagonal forms \cite{key-1}:

\begin{equation}
Q^{\rho}(A_{\mu}^{em})=g\left[(D^{\rho}\nu)\sin\theta+y_{\rho}\cos\theta\right]\label{Eq.7}\end{equation}

\begin{equation}
Q^{\rho}(Z_{\mu}^{i})=g\left[D_{k}^{\rho}-\nu_{k}(D^{\rho}\nu)(1-\cos\theta)-y_{\rho}\nu_{k}\sin\theta\right]\omega_{\cdot i}^{k\cdot}\label{Eq.8}\end{equation}

>From Eq. (7) one can write the following formula for the fundamental
multiplet ($y_{\rho}=0$):

\begin{equation}
g^{2}\sin^{2}\theta=2Tr(Q^{2})\label{Eq.9}\end{equation}

All the charges of the particles in a certain multiplet can be obtained
straightforwardly from this point just by taking into consideration
the multiplet's own representation $\rho$ with its hypercharge $y_{\rho}$value.

\section{The Exact Solution of a Special Class of 3-3-1 Models}

The general method can be accommodated to PPF model's fermionic content
(see Sec. 7 in Ref. \cite{key-1}) or to any other one's just by choosing
a particular set of versors in gWt which will lead to the desired
electric charges of the fundamental multiplet. In our paper, we ''conciliate''
the general method on one hand, with two particular ''no-exotic-electric-charge''
3-3-1 family models on the other hand, looking for their exact solution.
We focus our analysis on the boson mass issue. Since the fundamental
triplet in the class of models we intend to tackle is a quark triplet
belonging to the same representation of the local gauge group $q_{L}=\left(\begin{array}{ccc}
u & d & D\end{array}\right)_{L}^{T}\sim\left(\mathbf{3},\mathbf{3},0\right)$, we obtain just the two models C and D from Ref. \cite{key-9} when
we choose $\nu_{0}=0$, $\nu_{1}=0$ and $\nu_{2}=-1$ respectively.
The boson mass problem will be solved first by diagonalizing the mass
matrix and imposing certain conditions upon its eigenvectors. Then,
the charge computing in various representations will make the difference
between models.

\subsection{Higgs Sector}

The Higgs sector of any 3-3-1 model consists of three Higgs triplets
$\phi^{(1)}$, $\phi^{(2)}$ and $\phi^{(3)}$. The general parametrization
of the Higgs sector (which ensures a good non-degenerate boson mass
spectrum) with the property $Tr\eta^{2}=1-\eta_{0}^{2}$, reads in
our concrete case:

\begin{equation}
\eta^{2}=(1-\eta_{0}^{2})diag\left[\frac{1}{2}(a+b),1-a,\frac{1}{2}(a-b)\right]\label{Eq.10}\end{equation}
where, for the moment, $a$ and $b$ are arbitrary non-vanishing real
parameters. Obviously, $\eta_{0},a\in[0,1)$. 

Bearing in mind that for any model the above parametrization in the
general method should be added with the set $(g,\theta,\nu)$, we
opt for the following versors $\nu_{0}=0$, $\nu_{1}=0$, and $\nu_{2}=-1$
in order to obtain (using formula (7)) the correct electric charges
of the fundamental multiplet. We also establish from the very beginning
that the coupling constant of the model coincides with the first coupling
constant of the SM. The relation between the $\theta$ angle from
our parametrization and $\theta_{W}$ from SM arises from the constraint
$e=g\sin\theta_{W}$, the well-known condition in SM. Therefore, computing
Eq. (7) for the quark fundamental multiplet one gets $\sin\theta=\frac{2}{\sqrt{3}}$$\sin\theta_{W}$
which will be used in obtaining the eigenvalues of the mass matrix
(3). Thus, we can compute the elements of the mass matrices, according
to formulae (2) and (3).

\subsection{Boson Mass Spectrum}

For the non-diagonal bosons one obtains straightforwardly from Eq.
(2): $m_{W}^{2}=m^{2}a$, $m_{X}^{2}=m^{2}\left[1-\frac{1}{2}(a+b)\right]$
and $m_{Y}^{2}=m^{2}\left[1-\frac{1}{2}(a-b)\right]$, if we consider
$m^{2}=g^{2}\left\langle \phi\right\rangle ^{2}(1-\eta_{0}^{2})/4$
. 

For the diagonal (neutral) bosons the mass matrix will take the form

\begin{equation}
M^{2}=m^{2}\left|\begin{array}{ccc}
1-\frac{1}{2}a+\frac{1}{2}b &  & -\frac{1}{\sqrt{3-4s^{2}}}\left(1-\frac{3}{2}a-\frac{1}{2}b\right)\\
\\-\frac{1}{\sqrt{3-4s^{2}}}\left(1-\frac{3}{2}a-\frac{1}{2}b\right) &  & \frac{1}{3-4s^{2}}\left(1+\frac{3}{2}a-\frac{3}{2}b\right)\end{array}\right|\label{Eq.11}\end{equation}
where we have made the notation $\sin\theta_{W}=s$. 

Now, one has to accept that the old neutral boson $Z$ should be an
eigenvector of this mass matrix corresponding to the eigenvalue $m_{Z}^{2}=m_{W}^{2}/\cos^{2}\theta_{W}$
also established in the SM. That is, one computes $Det\left|M^{2}-m^{2}a/(1-s^{2})\right|=0$.
This leads to the following unique constraint upon the parameters
$b=a\tan^{2}\theta_{W}$. This result is very important since it shows
that a single parameter (let it be $a$) in the Higgs sector is enough
in order to exactly determine all the masses involved in the model.
Thus, a unique mass scale has been obtained. In this parametrization,
the masses of the gauge non-diagonal bosons become: $m_{W}^{2}=m^{2}a$,
$m_{X}^{2}=m^{2}\left(1-a/2\cos^{2}\theta_{W}\right)$ and $m_{Y}^{2}=m^{2}\left[1-a(1-\tan^{2}\theta_{W})/2\right]$.

Since $Tr(M^{2})=m_{Z}^{2}+m_{Z^{\prime}}^{2}$, the neutral diagonal
bosons will acquire the following masses:

\begin{equation}
m_{Z}^{2}=\frac{m^{2}a}{\cos^{2}\theta_{W}}\label{Eq.12}\end{equation}

\begin{equation}
m_{Z^{\prime}}^{2}=m^{2}\left[1+\frac{1}{3-4\sin^{2}\theta_{W}}-a\left(1+\frac{\tan^{2}\theta_{W}}{3-4\sin^{2}\theta_{W}}\right)\right]\label{Eq.13}\end{equation}

At this moment, the mass scale is just a matter of tuning the parameter
$a$ in accordance with the possible values for $<\phi>$. Thus, one
can play the game of recovering all the experimental values for the
bosons under consideration.

\subsection{The General Weinberg Transformation}

In our concrete 3-3-1 models, assuming the above versor choice, the
gWt reads

\[
A_{\mu}^{0}=A_{\mu}^{em}\cos\theta+\left(\omega_{\cdot1}^{2\cdot}Z_{\mu}^{1}+\omega_{\cdot2}^{2\cdot}Z_{\mu}^{2}\right)\sin\theta\]

\begin{equation}
A_{\mu}^{3}=\omega_{\cdot1}^{1\cdot}Z_{\mu}^{1}-\omega_{\cdot2}^{1\cdot}Z_{\mu}^{2}\label{Eq.14}\end{equation}

\[
A_{\mu}^{8}=-A_{\mu}^{em}\sin\theta+\left(\omega_{\cdot1}^{2\cdot}Z_{\mu}^{1}+\omega_{\cdot2}^{2\cdot}Z_{\mu}^{2}\right)\cos\theta\]
The transformation $\omega$ reduces here to a simple rotation of
angle $\theta^{\prime}$. Bearing in mind that the neutral boson mass
matrix becomes

\begin{equation}
M^{2}=m^{2}\left|\begin{array}{ccc}
\left[1-\frac{a}{2}\left(\frac{1-2s^{2}}{1-s^{2}}\right)\right] &  & -\frac{1}{\sqrt{3-4s^{2}}}\left[1-\frac{a}{2}\left(\frac{3-2s^{2}}{1-s^{2}}\right)\right]\\
\\-\frac{1}{\sqrt{3-4s^{2}}}\left[1-\frac{a}{2}\left(\frac{3-2s^{2}}{1-s^{2}}\right)\right] &  & \frac{1}{3-4s^{2}}\left[1+\frac{3}{2}a\left(\frac{1-2s^{2}}{1-s^{2}}\right)\right]\end{array}\right|\label{Eq.15}\end{equation}
one obtains

\begin{equation}
\omega=\frac{1}{2\sqrt{1-\sin^{2}\theta_{W}}}\left|\begin{array}{ccc}
\sqrt{3-4\sin^{2}\theta_{W}} &  & 1\\
\\-1 &  & \sqrt{3-4\sin^{2}\theta_{W}}\end{array}\right|\label{Eq.16}\end{equation}
which will lead (through gWt) to the charges of the particles. The
second neutral boson ($Z_{\mu}^{2}$ in our notation) will be the
Weinberg neutral boson while $Z_{\mu}^{1}$ is the new one of this
class of models.

\subsection{Charge Spectrum}

For the fundamental multiplet, the electric charge matrix can be expressed
in our parametrization - putting $y_{\rho}=0$ in Eq. (7) - as $Q(A_{\mu}^{em})=Diag(-e/3,-e/3,+2e/3)$.
The resulting matrix corresponds to both cases - models C and D in
Ref. \cite{key-9} - fitting the real values of the electric charges
of the known quarks. The neutral charges for the fundamental quark
triplet can be expressed from Eq.(8) by

\begin{equation}
Q(Z_{\mu})=\frac{e}{\sin2\theta_{W}}\left(-T_{3}+\frac{3-4\sin^{2}\theta_{W}}{\sqrt{3}}T_{8}\right)\label{Eq.17}\end{equation}

\begin{equation}
Q(Z_{\mu}^{\prime})=\frac{e\sqrt{3-4\sin^{2}\theta_{W}}}{\sin2\theta_{W}}\left(T_{3}+\frac{1}{\sqrt{3}}T_{8}\right)\label{Eq.18}\end{equation}

The electric charges of other multiplets (of the representations $\rho$)
can be obtained by adding to these expressions the hypercharges of
each multiplet accordingly. The neutral charges are then

\begin{equation}
Q^{\rho}(Z_{\mu})=\frac{e}{\sin2\theta_{W}}\left(\frac{2}{\sqrt{3}}\sin\theta_{W}\sqrt{3-4\sin^{2}\theta_{W}}y_{\rho}-T_{3}^{\rho}+\frac{3-4\sin^{2}\theta_{W}}{\sqrt{3}}T_{8}^{\rho}\right)\label{Eq.19}\end{equation}

\begin{equation}
Q^{\rho}(Z_{\mu}^{\prime})=\frac{e\sqrt{3-4\sin^{2}\theta_{W}}}{\sqrt{3}\sin2\theta_{W}}\left(\frac{2\sin\theta_{W}}{\sqrt{3-4\sin^{2}\theta_{W}}}y_{\rho}+\sqrt{3}T_{3}^{\rho}+T_{8}^{\rho}\right)\label{Eq.20}\end{equation}

Furthermore, if one computes Eqs. (7), (19), (20), and exploits the
substitutions $\alpha=\sqrt{3-4\sin^{2}\theta_{W}}$ and $\beta=\sin2\theta_{W}$,
one gets the values of the charges of the leptons in both models (Tables
1 and 2). We are surprised to discover by means of our method the
same values for all leptons as in the SM \cite{key-15}.

Based on techniques of the general method of exactly solving gauge
models with high symmetries we have obtained the exact solution of
two ''no-exotic-electric-charge'' 3-3-1 models. At this point they
are completely solved in the sense that no approximation is necesary
in order to establish the mass scale and to determine the coupling
coefficients. Moreover, we proved that the SM values are not spoiled
in this class of 3-3-1 model's exact solution.

\begin{table}

\caption{Charges of the particles in Model D}

\begin{tabular}{cccc}
\hline 
Charge\textbackslash{}Particle&
$e,\mu,\tau$&
$\nu_{e},\nu_{\mu},\nu_{\tau}$&
$N_{e}^{0},N_{\mu}^{0},N_{\tau}^{0}$\tabularnewline
\hline
\hline 
$Q(A^{em})$&
$-e$&
$0$&
$0$\tabularnewline
&
&
&
\tabularnewline
$Q(Z)$&
$-\frac{e}{\beta}(2\sin^{2}\theta_{W}-1)$&
$-\frac{e}{\beta}$&
$0$\tabularnewline
&
&
&
\tabularnewline
$Q(Z^{\prime})$&
$-\frac{e\alpha}{3\beta}(\frac{2\sin^{2}\theta_{W}}{\alpha^{2}}-1)$&
$-\frac{e\alpha}{3\beta}(\frac{2\sin^{2}\theta_{W}}{\alpha^{2}}-1)$&
$-\frac{2e\alpha}{3\beta}(\frac{\sin^{2}\theta_{W}}{\alpha^{2}}+1)$\tabularnewline
\hline
\hline 
&
&
&
\tabularnewline
\end{tabular}
\end{table}

\begin{table}

\caption{Charges of the particles in Model C}

\begin{tabular}{cccc}
\hline 
Charge\textbackslash{}Particle&
$e,\mu,\tau$&
$\nu_{e},\nu_{\mu},\nu_{\tau}$&
$E,T,M$\tabularnewline
\hline
\hline 
$Q(A^{em})$&
$-e$&
$0$&
$-e$\tabularnewline
&
&
&
\tabularnewline
$Q(Z)$&
$-\frac{e}{\beta}(2\sin^{2}\theta_{W}-1)$&
$-\frac{e}{\beta}$&
$-\frac{e}{\beta}2\sin^{2}\theta_{W}$\tabularnewline
&
&
&
\tabularnewline
$Q(Z^{\prime})$&
$-\frac{e\alpha}{3\beta}(\frac{4\sin^{2}\theta_{W}}{\alpha^{2}}+1)$&
$-\frac{e\alpha}{3\beta}(\frac{4\sin^{2}\theta_{W}}{\alpha^{2}}+1)$&
$-\frac{2e\alpha}{3\beta}(\frac{2\sin^{2}\theta_{W}}{\alpha^{2}}-1)$\tabularnewline
\hline
\hline 
&
&
&
\tabularnewline
\end{tabular}
\end{table}

\subsection{Fermion Sector}

In the following, we focus on the particular fermion content of Model
D. It consists of three fermion generations: three lepton generations
with the same representation and three quark generations that obey
different representations with respect to the gauge group of the model.
Lepton triplets are color singlets and quark triplets are color triplets.
Lepton families are\begin{equation}
\begin{array}{ccccc}
f_{\alpha L}=\left|\begin{array}{c}
e_{\alpha}\\
\nu_{\alpha}\\
\nu_{\alpha}^{c}\end{array}\right|_{L}\sim(\mathbf{1,3^{\mathbf{\mathbf{*}}}},-1/3) &  &  &  & e_{\alpha}^{c}\sim(\mathbf{1},\mathbf{1},1)\end{array}\label{Eq.21}\end{equation}
where $e_{\alpha}=e,\mu,\tau$. We mean by superscript $c$ the charge
conjugation (for details see Appendix B in \cite{key-1}) and by subscript
$L$ the chiral left-handed component. Evidently, the third component
of the triplet is now considered as the right-handed component of
neutrino field. This assumption does not alter the previous results
and it is most likely since the right-handed neutrino does not couple
to the SM neutral gauge boson (see Table 1). Therefore, it had no
reason to be part of the SM.

The \emph{}quarks come (according to the anomaly cancellation requirement)
in three distinct generations as follows: 

\begin{equation}
\begin{array}{ccc}
Q_{iL}=\left|\begin{array}{c}
u_{i}\\
d_{i}\\
D_{i}\end{array}\right|_{L}\sim(\mathbf{3,3^{*}},0) &  & Q=\left|\begin{array}{c}
d\\
u\\
U\end{array}\right|_{L}\sim(\mathbf{3},\mathbf{3}^{\mathbf{*}},1/3)\end{array}\label{Eq.22}\end{equation}

\begin{equation}
\begin{array}{ccc}
(d_{L})^{c},(d_{iL})^{c}\sim(\mathbf{3^{\mathbf{*}}},\mathbf{1},1/3) &  & (u_{L})^{c},(u_{iL})^{c}\sim(\mathbf{3^{\mathbf{*}}},\mathbf{1},-2/3)\end{array}\label{Eq.23}\end{equation}

\begin{equation}
\begin{array}{ccccccccc}
(U_{L})^{c}\sim(\mathbf{3^{\mathbf{*}},1},-2/3) &  &  &  &  &  &  &  & (D_{iL})^{c}\sim(\mathbf{3^{\mathbf{*}},1},1/3)\end{array}\label{Eq.24}\end{equation}
with $i=1,2$.

\subsection{Yukawa Sector}

For quarks and leptons - except for neutrinos - the traditional Yukawa
couplings seem to be sufficent in order to get their desired masses
\cite{key-4} - \cite{key-8}. 

The lepton families in 3-3-1 models under consideration here acquire
their masses through the following couplings \cite{key-4} - \cite{key-8}:

\begin{equation}
\mathcal{-L}_{Y}=G_{\alpha\alpha}{}\bar{f}_{\alpha L}\rho e_{\alpha L}^{c}+G_{\alpha\beta}\varepsilon^{ijk}\left(\bar{f}_{\alpha L}\right)_{i}\left(f_{\beta L}^{c}\right)_{j}\left(\rho^{*}\right)_{k}+H.c.\label{Eq.25}\end{equation}
The Higgs triplets should transform as $\chi,\eta\sim\left(\mathbf{1,3^{\mathbf{*}}},-1/3\right)$
and $\rho\sim\left(\mathbf{1,3^{\mathbf{*}}},2/3\right)$ in order
to match the gauge invariance conditions when they couple to fermion
fields in Yukawa terms Eq. (25). Their electric charge assignment
leads to $\chi=\left(\chi_{-},\chi_{0},\chi_{0}\right)^{T}$, $\eta=\left(\begin{array}{ccc}
\eta_{-}, & \eta_{0}, & \eta_{0}\end{array}\right)^{T}$ and $\rho=\left(\begin{array}{ccc}
\rho_{0} & ,\rho_{+}, & \rho_{+}\end{array}\right)^{T}$. These scalar fields acquire, when looking for the local minimum
of the scalar potential in Eq. (1), the following VEVs: $<\chi>=\left(\begin{array}{ccc}
0, & 0, & V_{\chi}\end{array}\right)^{T}$, $<\eta>=\left(\begin{array}{ccc}
0, & V_{\eta}, & 0\end{array}\right)^{T}$ and $<\rho>=\left(\begin{array}{ccc}
V_{\rho}, & 0, & 0\end{array}\right)^{T}$ which generate masses of the quarks and charged leptons as in \cite{key-4}
- \cite{key-8}. 

According to a traditional Dirac Ld put in the pure left form (see
Appendix B in Ref. \cite{key-1}), one can identify the mass of the
lepton as

\begin{equation}
m(e_{\alpha})=G_{\alpha\alpha}<\phi^{(\rho)}>\label{Eq.26}\end{equation}
In our method each mass source corresponds to a certain parameter
which multiplies the same VEV $<\phi>$. The exact solution shown
above demands the following relation between the two parameters $b=a\tan^{2}\theta_{W}$.
Thus, $\eta$ becomes a one-parameter matrix $\eta^{2}=\left(1-\eta_{0}^{2}\right)diag\left[a/2\cos^{2}\theta_{W},1-a,a(1-\tan^{2}\theta_{W})/2\right]$.
Under these circumstances, for each $a$ one can estimate the alignment
of VEVs just by bijective mapping $(\chi,\rho,\eta)\rightarrow(1,2,3)$
while keeping unchanged the order in the parameter matrix $\eta$.
The charged lepton mass can have, with respect to the parameter mapping,
the following possible values: 

\begin{itemize}
\item $m(e_{\alpha})=G_{\alpha\alpha}\sqrt{1-a}<\phi>$ (case I) 
\item $m(e_{\alpha})=G_{\alpha\alpha}\sqrt{a(1-\tan^{2}\theta_{W})/2}<\phi>$
(case II) 
\item $m(e_{\alpha})=G_{\alpha\alpha}\sqrt{a/2\cos^{2}\theta_{W}}<\phi>$
(case III). 
\end{itemize}
Up to this stage each of these cases has two subcases for the choice
of the remaining two scalar triplets.

According to the gauge invariance requirement, the Yukawa coupling
terms for quark families are the same as those in Refs. \cite{key-4}
- \cite{key-8} and they supply a good quark mass spectrum within
the framework of exactly solving method as well. 

If one introduces the more restrictive requirement of getting neutrino
mass, then only two out of the six possible cases can be taken into
consideration (as we will see in Sec. 4).

\section{Neutrino Mass}

It has been considered for a long time that the neutrino is a massless
elementary particle that played a crucial role in the weak sector
of the SM. Recently, certain evidences regarding the new phenomenon
of ''neutrino oscillations'' were found in Super-Kamiokande \cite{key-16}
- \cite{key-18}, SNO \cite{key-19} - \cite{key-21}, KamLAND \cite{key-22},
K2K \cite{key-23}, and other neutrino experiments \cite{key-24}
- \cite{key-30}. These suggest a small but non-zero neutrino mass,
although the nature of massive neutrinos remains an open issue. Are
they Dirac or Majorana particles? Definitely, they are the signature
of new physics beyond SM. Therefore, some attempts to embed neutrino
mass in various extensions of the SM seem justified. In 3-3-1 models
these attempts gave promising results by using distinct strategies:
radiative mechanisms (1 - and 2 - loop radiative corrections \cite{key-31}
- \cite{key-33}), the Higgs triplet method in exotic models \cite{key-34},
or various see-saw mechanisms \cite{key-35} - \cite{key-37}.

We propose a different method in order to provide neutrino masses
using the exact solution presented above for a particular 3-3-1 model
(namely model D). Neutrinos are considered here as Majorana fields
and they enter naturally into the mass mixing matrix straightforwardly
from the Yukawa terms after SSB took place.

\subsection{Neutrino Mass Matrix }

In order to obtain masses for neutrinos involved in model D, one should
introduce in Eq. (25) a new term $G_{\alpha\beta}{}_{L}\bar{f}_{\alpha L}Sf_{\beta L}^{c}+H.c.$,
with $S\sim(\mathbf{1},\mathbf{6},-2/3)$ a symmetric matrix constructed
out of a sum of tensor products among certain Higgs multiplets in
the manner $S=\phi^{-1}\left(\phi^{(\eta)}\otimes\phi^{(\chi)}+\phi^{(\chi)}\otimes\phi^{(\eta)}\right)$.
It looks like: 

\begin{equation}
S=\phi^{-1}\left(\begin{array}{ccc}
\phi_{1}^{(\eta)}\phi_{1}^{(\chi)}+\phi_{1}^{(\chi)}\phi_{1}^{(\eta)} & \phi_{2}^{(\eta)}\phi_{1}^{(\chi)}+\phi_{2}^{(\chi)}\phi_{1}^{(\eta)} & \phi_{3}^{(\eta)}\phi_{1}^{(\chi)}+\phi_{3}^{(\chi)}\phi_{1}^{(\eta)}\\
\phi_{1}^{(\eta)}\phi_{2}^{(\chi)}+\phi_{1}^{(\chi)}\phi_{2}^{(\eta)} & \phi_{2}^{(\eta)}\phi_{2}^{(\chi)}+\phi_{2}^{(\chi)}\phi_{2}^{(\eta)} & \phi_{3}^{(\eta)}\phi_{2}^{(\chi)}+\phi_{3}^{(\chi)}\phi_{2}^{(\eta)}\\
\phi_{1}^{(\eta)}\phi_{3}^{(\chi)}+\phi_{1}^{(\chi)}\phi_{3}^{(\eta)} & \phi_{2}^{(\eta)}\phi_{3}^{(\chi)}+\phi_{2}^{(\chi)}\phi_{3}^{(\eta)} & \phi_{3}^{(\eta)}\phi_{3}^{(\chi)}+\phi_{3}^{(\chi)}\phi_{3}^{(\eta)}\end{array}\right)\label{Eq.27}\end{equation}
and after SSB only $\phi_{2}^{(\eta)}$ and $\phi_{3}^{(\chi)}$ will
survive gauge fixing. Therfore, only the positions where they meet
(23 and 32 in $<S>$) will be non-zero.

\begin{equation}
<S>=\frac{1}{<\phi>}\left(\begin{array}{ccc}
0 & 0 & 0\\
0 & 0 & <\phi^{(\chi)}><\phi^{(\eta)}>\\
0 & <\phi^{(\eta)}><\phi^{(\chi)}> & 0\end{array}\right)\label{Eq.28}\end{equation}

The neutrino mass matrix takes then the form:

\begin{equation}
M=4\left|\begin{array}{ccc}
A & D & L\\
E & B & F\\
K & G & C\end{array}\right|\frac{<\phi^{(\eta)}><\phi^{(\chi)}>}{<\phi>}\label{Eq.29}\end{equation}
with the coupling constants $A=G_{ee}$, $B=G_{\mu\mu}$, $C=G_{\tau\tau}$,
$D=G_{e\mu}$, $E=G_{\mu e}$, $F=G_{\mu\tau}$ , $G=G_{\tau\mu}$,
$L=G_{e\tau}$, $K=G_{\tau e}$. We notice that for the Majorana case,
matrix $M$ is a symmetric one, that is $D=E$, $F=G$, $L=K$.

At this point there are three possible situations (depending on the
VEV alignment choice) which have to be separately analyzed, all of
them containing the sole parameter $a$. Some of them will be ruled
out by certain restrictive conditions imposed by phenomenological
reasons.

\subsection{Neutrino Mixing}

The physical neutrino mass issue can be addressed if we consider first
neutrino mixing (for details see the excellent reviews of Bilenky
\cite{key-38} - \cite{key-40} or Mohapatra \cite{key-41} and references
therein). The unitary mixing matrix $U$ (with $U^{+}U=1$) links
the gauge-flavor basis to the physical basis of massive neutrinos
in the manner:

\begin{equation}
\nu_{\alpha L}(x)=\sum_{i=1}^{3}U_{\alpha i}\nu_{iL}(x)\label{Eq.30}\end{equation}
where $\alpha=e,\mu,\nu$ (corresponding to neutrino gauge eigenstates),
and $i=1,2,3$ (corresponding to massive physical neutrinos with masses
$m_{i}$). In the following, we consider all neutrinos as Majorana
fields, \emph{i.e.} $\nu_{iL}^{c}(x)=\nu_{iL}(x)$. This is very plausible,
since neutrinoless double $\beta$-decay is still under experimental
fire. The mass term in the Yukawa sector yields: 

\begin{equation}
\mathcal{-L}_{\nu}^{mass}=\frac{1}{2}\bar{\nu}_{\alpha L}M_{\alpha\beta}\nu_{\beta L}^{c}+H.c\label{Eq.31}\end{equation}
The mixing matrix $U$ that diagonalizes the mass matrix $U^{T}MU=m_{ij}\delta_{j}$
has in the standard parametrization the form: 

\begin{equation}
U=\left|\begin{array}{ccc}
c_{2}c_{3} & s_{2}c_{3} & s_{3}e^{-i\delta}\\
-s_{2}c_{1}-c_{2}s_{1}s_{3}e^{i\delta} & c_{1}c_{2}-s_{2}s_{3}s_{1}e^{i\delta} & c_{3}s_{1}\\
s_{2}s_{1}-c_{2}c_{1}s_{3}e^{i\delta} & -s_{1}c_{2}-s_{2}s_{3}c_{1}e^{i\delta} & c_{3}c_{1}\end{array}\right|\label{Eq.32}\end{equation}
where we made the substitutions $\sin\theta_{23}=s_{1}$, $\sin\theta_{12}=s_{2}$,
$\sin\theta_{13}=s_{3}$, $\cos\theta_{23}=c_{1}$, $\cos\theta_{12}=c_{2}$,
$\cos\theta_{13}=c_{3}$ for the mixing angles, and $\delta$ is the
CP phase.

\subsection{Restrictions on Parameter \textmd{$a$}}

Assuming that the new neutral boson $Z^{\prime}$ has to be heavier
than the Weinberg neutral boson $Z$, when comparing the masses of
the two bosons (formulae (12) - (13)) one obtains the condition:

\begin{equation}
a<\frac{2(1-\sin^{2}\theta_{W})}{3-2\sin^{2}\theta_{W}}\label{Eq.33}\end{equation}
Now, if we take $\sin^{2}\theta_{W}=0.23113\pm0.00015$ \cite{key-14}
we get $a<0.60594\pm0.00005$.  The first remark is that parameter
$a$ is upper bounded in all the three cases under consideration.

Important information comes also from the trace of matrix $M$ when
combining Eqs. (26) and (29):

\begin{equation}
TrM=4m(\tau)\left[1+\frac{m(\mu)}{m(\tau)}+\frac{m(e)}{m(\tau)}\right]\frac{<\phi^{(\chi)}><\phi^{(\eta)}>}{<\phi^{(\rho)}><\phi>}{}\label{Eq.34}\end{equation}
Bearing in mind that, at the same time, $TrM=\sum_{i}m_{i}$ and phenomenological
values $m_{i}$ of neutrino masses are severely limited to few eVs,
one remains with Case I only. Case II and Case III are ruled out since
both they give $TrM\sim\sqrt{1-a}$ . These traces reach their minimum
values when parameter $a$ reaches its maximum allowed value. Thus,
these cases are not compatible with the right order of magnitude for
neutrino masses that can be provided only by very small values of
parameter $a$ in Case I. Therefore, a good solution is:

\begin{equation}
\sum_{i}m_{i}\simeq m(\tau)\left(\frac{2a}{\sqrt{1-a}}\right)\frac{\sqrt{1-2\sin^{2}\theta_{W}}}{\cos^{2}\theta_{W}}\label{Eq.35}\end{equation}
where we neglected the small ratios $m(\mu)/m(\tau)\sim0.05$ and
$m(e)/m(\tau)\sim0.0002$ in Eq. (34). 

Now, if we consider as valid the values offered by Tritium $\beta-$decay
experiments Troitsk \cite{key-42} and Mainz \cite{key-43} - \cite{key-44}
which consider that $m_{\beta}\leq2.3$ eV, then parameter $a$ has
to be in the range $a<0.678\cdot10^{-9}$. This implies for VEV $<\phi>\sim0.95\cdot10^{6}$TeV
and the resulting masses for the new bosons: $m_{X}\simeq m_{Y}>3.1\cdot10^{3}$
TeV and $m_{Z^{\prime}}>3.7\cdot10^{3}$ TeV. 

Based on cosmological data \cite{key-45}, the upper limit is even
more restricted $\sum_{i}m_{i}<(0.4-1.7)$ eV, which leads to $a<(0.118-0.501)\cdot10^{-9}$.
Consequently, bosons will acquire masses: $m_{X}\simeq m_{Y}>(3.6-7.5)\cdot10^{3}$
TeV and $m_{Z^{\prime}}>(4.4-9.1)\cdot10^{3}$ TeV at a breaking scale
$<\phi>\sim(1.1-2.3)\cdot10^{6}$ TeV. 

While the ''old'' bosons remain at their SM values, the ''new''
ones now have to become very massive. This is the price paid in order
to have good phenomenological results for all the particles in this
3-3-1 model, using only one free parameter. Notwithstanding, we note
that such massive bosons do not contradict the possible mass evaluation
communicated in Ref. \cite{key-14} where they are accepted as $m_{X}\simeq m_{Y}>800$
GeV and $m_{Z^{\prime}}>1500$ GeV. Therefore, one can consider that
neutrino masses could arise (avoiding the see-saw prescriptions, the
method of 1 - and 2 - loop radiative mechanisms or any other approximations),
only as a consequence of a very large breaking scale of the model
and accompanied by very massive bosons.

\subsection{Mass Squared Differences}

For physical neutrinos, mass squared differences - which are experimentally
accesible - are defined as $\Delta m_{ij}^{2}=m_{j}^{2}-m_{i}^{2}$.
Their right order of magnitude can be obtained for $\Delta m_{12}^{2}\leq8\cdot10^{-5}$
eV$^{2}$ from solar and KamLAND data \cite{key-19} - \cite{key-22}
and for $\Delta m_{23}^{2}\leq2\cdot10^{-3}$ eV$^{2}$ from Super
Kamiokande atmospheric data \cite{key-16} - \cite{key-18}. 

Considering that in Eq. (29) the coupling constants act as variables,
the diagonalization of the matrix $M$ is equivalent to a system of
9 linear equations with 12 variables (6 linear equations with 9 variables,
if $M$ is symmeric - Majorana case) which leads to the following
solution for the physical neutrino masses:

\begin{equation}
m_{i}=f_{i}(A,B,C)\frac{<\phi^{(\chi)}><\phi^{(\eta)}>}{<\phi>}\label{Eq.36}\end{equation}
for $i=1,2,3$. They can be put in a more explicit form of the Case
I

\begin{equation}
m_{i}=F_{i}\left[\frac{m(\mu)}{m(e)},\frac{m(\tau)}{m(e)},\theta_{12},\theta_{13},\theta_{23}\right]\left(\frac{a}{\sqrt{1-a}}\right)\frac{\sqrt{1-2\sin^{2}\theta_{W}}}{2\cos^{2}\theta_{W}}m(e)\label{Eq.37}\end{equation}
where $F_{i}$ are linear combinations of mass ratios:

\begin{equation}
F_{i}\left[\frac{m(\mu)}{m(e)},\frac{m(\tau)}{m(e)},\theta_{12},\theta_{13},\theta_{23}\right]=\alpha_{i}+\beta_{i}\frac{m(\mu)}{m(e)}+\gamma_{i}\frac{m(\tau)}{m(e)}\label{Eq.38}\end{equation}
Coefficients $\alpha_{i}$, $\beta_{i}$, $\gamma_{i}$ are analytical
functions depending on mixing angles and they result from solving
the linear system of diagonalization. 

The mass squared differences are now:

\begin{equation}
\Delta m_{ij}^{2}=(F_{j}^{2}-F_{i}^{2})\left(\frac{a}{\sqrt{1-a}}\right)^{2}\frac{(1-2\sin^{2}\theta_{W})}{4\cos^{4}\theta_{W}}m^{2}(e)\label{Eq.39}\end{equation}
A more detailed analysis of the resulting values for mass squared
differences (when LMA conditions are embedded in our method) will
be presented elsewhere. Here we restrict ourselves to observing only
that for $a\sim10^{-9}-10^{-10}$ the necessary order of magnitude
for $(F_{j}^{2}-F_{i}^{2})$ has to be $\sim10^{6}$ in order to fit
experimental data. This is quite plausible, assuming that in Eq. (38)
the leading (larger) term is $\gamma_{i}m(\tau)/m(e)$ and its order
of magnitude is at most $\sim10^{3}$.

\section{Concluding Remarks}

We have presented an original method of exploring neutrino mass on
theoretical grounds, within the framework of the exact solution of
a particular 3-3-1 gauge model. The unusual coupling among the lepton
multiplets (tied together by tensor-like products among Higgs multiplets)
led to an interesting solution in accordance with actual phenomenological
data. 

Concerning the general method applied here, we observe that the ''geometrization''
of the Higgs sector requires only one parameter (since $\eta_{0}$
can be incorporated in the VEV $<\phi>$). In order to investigate
mass spectrum, one has to tune the sole parameter $a$. For instance,
if $a\simeq1$ then the mass scale will be upper bounded by the Weinberg
neutral boson's mass and the new neutral boson will gain a mass of
about $m_{Z^{\prime}}\simeq m_{Z}/2$. This situation is definitely
ruled out by phenomenological reasons. If the more plausible $a\simeq0$
occurs, then the upper limit of the mass scale will be given by the
new neutral boson which could stand, in principle, at any level coming
from VEV $<\phi>$. The value can be accommodated to fit the known
masses of the ''old'' bosons in the model. If one wants to keep
the $m_{Z^{\prime}}>m_{Z}$ condition, then the parameter must be
$a<0.6$. More severe restrictions on this parameter which require
the range $a\sim10^{-9}$ come from neutrino data. 

Inputting from the beginning the set of parameters $(g,\theta,\nu)$,
one observes for models under consideration here that the ''old''
charges of the particles reproduce through our method their values
from SM, while the ''new'' charge presents indistinct values for
the SM leptons. This result makes obvious the fact that, with respect
to this new boson, the SM leptons present the same behavior and that
could be the sign of the new physics hidden by this new boson at energies
above the TeV scale. At the same time, the charges are independent
from the boson masses, also as in the SM. One can observe for Model
D that the new neutral particles (right-handed neutrinos) do not couple
to $Z$. This could explain why the right-handed neutrinos are not
manifest in SM and - as a consequence - why neutrino mass had to be
identical to zero in SM. 

Assuming that parameter $a<0.6$ one can evaluate a unique mass scale
for all fermions and gauge bosons in the model, encountering three
possible cases for the alignment of VEVs $<\phi^{(\eta)}>$, $<\phi^{(\rho)}>$,
$<\phi^{(\chi)}>$. In order to keep our method flexible we eliminated
cases II and III (Sec. 4) - where ''the smaller the parameter, the
larger the neutrino mass'' - for they give an unacceptable order
of magnitude for the neutrino mass when $a\rightarrow0$. Only case
I allows $m(\nu_{i})\ll m(e)$ and therefore it seems to be the most
likely. In addition, the upper limit on the values of parameter $a$
offers a lower limit on the VEV $<\phi>$. When comparing the mass
of the Weinberg neutral boson $Z$ obtained at a scale $<\phi_{0}>\sim174$
GeV in the SM to the mass of the same boson obtained in our 3-3-1
model, one has to impose that these ones be identical. This requirement
leads to the following condition $<\phi>\geq316$ GeV which is also
plausible, leading to the required larger values for the breaking
scale in order to keep consistency with neutrino mass phenomenology. 

In conclusion, we consider that the exact solution of the 3-3-1 models
without exotic electric charges presented above can act as a viable
theoretical background for the electro-weak phenomenology. We proved
that it can fit all the available experimental data and predict the
existence of very heavy new bosons, using only one free parameter.
These are the basic requirements of a good electro-weak theory that
in addition includes (at a very large breaking scale) a suitable mechanism
which allows neutrinos to aquire their Majorana masses.

\section*{Acknowledgement}

\paragraph*{\textmd{The author would like to thank I.I. Cot\u{a}escu for fruitful
discussions.}}

\end{document}